\documentclass[intlimits,twoside,a4paper]{article}

\usepackage{graphicx}
\usepackage[T2A]{fontenc}
\usepackage[cp1251]{inputenc}
\usepackage{amsmath}
\usepackage{siunitx}
\usepackage[eqsecnum]{cmpj2}

\issue{2013}{16}{4}{43002}
\doinumber{10.5488/CMP.16.43002}

\title[Potential of electrorheological fluids]%
{The origin of interparticle potential of electrorheological fluids%
\thanks{We dedicate this paper to Myroslav Holovko in honour of his valuable contribution to the field of statistical mechanics of fluids.}
}
\author[D.~Boda, M.~Valisk\'o, I.~Szalai]{D. Boda\refaddr{label1}\thanks{E-mail: boda@almos.vein.hu}\,, M. Valisk\'o\refaddr{label1}, I. Szalai\refaddr{label2}
}

\authorcopyright{D.~Boda, M.~Valisk\'o, I.~Szalai, 2013}

\addresses{
 \addr{label1} Department of Physical Chemistry, University of Pannonia, P. O. Box 158, Veszpr\'em, Hungary
 \addr{label2} Institute of Physics and Mechatronics, University of Pannonia, P. O. Box 158, Veszpr\'em, Hungary
}

\date{Received July 29, 2013}

\begin{document}

\maketitle

\begin{abstract}
The particles of electrorheological fluids can be modelled as dielectric spheres (DS) immersed in a continuum dielectric.
When an external field is applied, polarization charges are induced on the surfaces of the spheres and can be represented as point dipoles placed in the centres of the spheres.
When the DSs are close to each other, the induced charge distributions are distorted by the electric field of the neighbouring DSs.
This is the origin of the interaction potential between the DSs.
The calculation of this energy is very time consuming, therefore, the DS model cannot be used in molecular simulations.
In this paper, we show that the interaction between the point dipoles appropriately approximates the interaction of DSs.
The polarizable point dipole model provides better results, but this model is not pair-wise additive, so it is not that practical in particle simulations.

\pacs 07.05.Tp, 47.65.Gx, 83.80.Gv, 34.20.Gj, 31.30.jn, 77.84.-s
\end{abstract}

\section{Introduction}

Electrorheological (ER) fluids are suspensions of fine non-conducting solid particles (up to 50~{\textmu}m diameter) in an electrically insulating liquid.
The dielectric permittivity of dispersed particles is usually higher than that of the carrier liquid \cite{gast1}.
The rheological properties of ER fluids are controllable by the application of an electric field \cite{sheng1}.
The apparent viscosity of an ER fluid increases abruptly by the application of a strong electric field of the order of kilovolts per millimeter
[silica particles $(\rm{SiO}_2)$ dispersed in silicone oil is a typical electrorheological fluid].
The electric field causes a reversible change in the viscosity.
The increase of the apparent viscosity is caused by the chain and column formation of the grains carrying electric dipole moments induced by the external field \cite{gast1,hongru1}.
Due to the electric-field-induced aggregation, the dielectric properties of the ER fluids are also changed \cite{horvath1}.

Electrorheological particles are beyond the molecular scale, therefore, their modelling necessarily includes some coarse graining.
Coarse-graining means that the many-atom system is modelled by averaging certain degrees of freedom into a response function.
The atoms of the ER grains are not modelled explicitly, instead, their dielectric response is taken into account by a dielectric continuum characterized by a certain dielectric constant.
The same is performed for the carrier liquid, but with a different dielectric constant.

Therefore, in this coarse-grained but realistic picture, the particle is modelled as a dielectric sphere (DS) immersed in a dielectric continuum.
The two interacting particles carry three-dimensional surface charge distributions on their surfaces induced by the external electric field and the electric field exerted by the other particles.
Computation of this induced charge distribution is a non-trivial and time consuming process. Therefore, this model is not feasible in computer simulations.
Thus, a more simplified model is needed for simulations, possibly a model with pair-wise additive interactions.
Such a model for the particles of ER fluids is a sphere carrying a point dipole in its center (we refer to this model as DD).
The dipoles are induced by an external electric field $\mathbf{E}$, all being aligned in its direction.
The lowest energy configuration of two dipoles are the head-to-tail position where the two dipoles are aligned in the same direction along the same line.
Therefore, as we have mentioned, the particles in electrorheological fluids form chains in the presence of an external field.
This chain formation is responsible for the increase of viscosity when an electric field is applied.

Simulation study of chain formation of ER fluids is based on different models.
Klingenberg et al. \cite{klingenberg1} used the interaction of single point dipoles with hard core repulsion while Bonnecaze and Brady \cite{bonnecaze1} used a sophisticated polarization model for the dipole-dipole interaction.
Chain formation was also found in fluids of particles carrying strong permanent dipoles \cite{weis1, szalai1}.
The orientation of these dipoles, however, was not restricted.
This distinction makes that case different from ER fluids, where the dipoles are induced dipoles with directions fixed by the external field.

The ideal point dipole is clearly an approximation to the charge distribution of the more realistic DS model.
The DD model ignores the fact that the spheres are polarized not only by the external field but also by the other spheres.
This effect can be taken into account by the polarizable dipole (PD) model, which is a step further but still an approximation to the DS model.
The PD model is not pair-wise additive, but still feasible in simulations because we have to compute the potentials/fields only at the particle centers rather than on the whole particle surfaces.

The PD model was also used in our simulation study \cite{valisko_cm} of the correction to the Clausius-Mosotti equation  describing the dielectric constant of non-polar fluids.
In this work, the non-polar particles were also polarized by a uniform external field.
The analogy with electrorheological fluids is unmistakable.

In this paper, we study the DS model and compare its energetics with that given by the DD and PD models by computing the interaction energy between two spheres using all the three models.
We conclude that the PD model is an excellent, while the DD model is a reliable approximation to the DS model.

\section{The dielectric sphere model}

The dielectric constant inside the sphere, $\epsilon_i$, is different from that outside the sphere, $\epsilon_\mathrm{w}$, (the subscript $i$ refers to an ER particle species here). Then, a dielectric boundary is formed at the surface of the sphere. The external electric field induces a surface charge distribution on this boundary. Since there is no free charge inside the sphere (it is neutral), the net induced charge is zero according to Gauss's law. Thus, the separation of charge on the surface of the sphere corresponds to a dipole-like distribution. Using basic electrostatics in terms of Legendre polynomials\ \cite{jackson, griffiths}, the dipole moment can be computed as
\begin{equation}
 \mathbf{p}_{i}=\dfrac{\epsilon_{i}-\epsilon_{\mathrm{w}}}{\epsilon_{i}+2\epsilon_{\mathrm{w}}}a_{i}^{3}\textbf{E} = \alpha_{i} \textbf{E},
\label{dipmom}
\end{equation}
 where $a_{i}$ is the radius of the sphere and
\begin{equation}
 \alpha_{i}=\dfrac{\epsilon_{i}-\epsilon_{\mathrm{w}}}{\epsilon_{i}+2\epsilon_{\mathrm{w}}}a_{i}^{3}
\end{equation}
is the polarizability of the sphere. When the external field is uniform, this is an exact solution: the electric field outside the sphere is equal to the electric field of a point dipole in the center of the sphere.

These charges on the surface of the spheres are not free charges, but rather they are bound (induced) charges. This means that they do not get there from some external circuit, because they have always been there. Their appearance is due to polarization: an electric field separates positive and negative charges of the dielectric. When we compute the energy of a macroscopic dielectric system, the interaction between bound charges does not appear in the formulation. The total energy of the system is the work done against electric field as we bring in the free charges from infinity, namely the work needed to build up the charge distribution in a charge-up process. If we denote the distribution of free charges in the system by $q(\mathbf{r})$, and that of the induced charges by $h(\mathbf{r})$, then this work is computed as follows:
\begin{equation}
 W=\dfrac{1}{2}\int q(\mathbf{r}) \Psi(\mathbf{r}) \rd V ,
\label{eq:W}
\end{equation}
where $\Psi(\mathbf{r})=\Psi_{q}(\mathbf{r})+\Psi_{h}(\mathbf{r})$ contains the potentials produced both by the free charges
\begin{equation}
\Psi_{q}(\mathbf{r})=\int \dfrac{q(\mathbf{r}')}{|\mathbf{r}-\mathbf{r}'|}\rd V'
\label{e:2.4}
\end{equation}
and by the induced (bound) charges
\begin{equation}
\Psi_{h}(\mathbf{r})=\int \dfrac{h(\mathbf{r}')}{|\mathbf{r}-\mathbf{r}'|}\rd V'.
\label{e:2.5}
\end{equation}
Equations use Gaussian units. If the free charges are point charges, the integral in equation~\eqref{e:2.5} becomes a sum. If the dielectric boundary is sharp, the induced charge distribution is a surface charge, so the integral becomes a surface integral. In an experiment, the external electric field is produced by surface charges $\sigma_{1}$  and $\sigma_{2}$ on the plates of a capacitor. In this case, the integral in equation~\eqref{e:2.4} also becomes a surface integral.

In the above equations, the free charge --- free charge interaction and the free charge --- induced charge interaction appear. The induced charge --- induced charge interaction is missing. If we write up the energy as the sum of the interactions between all charges (including free and induced charges), an additional term has to be added. This is the work involved in stretching the dielectric molecules, namely, the work necessary to polarize the dielectric.
This work is equal in magnitude and opposite in sign to the induced charge --- induced charge interaction, so they cancel.
That is why only the energy of the free charges appears in the equation for the total electrostatic energy of the system [see equation~\eqref{eq:W}].

This result is counter-intuitive: one might think that we have to take into account the interaction between all charges in the system. Moreover, it is counter-intuitive regarding the dipoles of the ER particles. In the DD approach, the interaction energy is computed from the interaction of the dipoles. The dipoles are interpretations of the induced charges. This stands a paradox: why do we take into account the direct interaction between the charge distributions of the particles in the DD approach and why do not we take it into account in the DS approach? A goal of this paper is to resolve the apparent conflict between the two approaches. We will shed some light on the mechanism of the interaction between the ER particles.

The induced charge is calculated by the Induced Charge Computation (ICC) method\ \cite{boda-pre-69,Boda_NATO,boda-jcp-125-2006}.
This is a boundary element method where the dielectric boundary surface is divided into surface elements. Poisson's equation is transformed into an integral equation where the unknown variable is the discretized induced charge treated as constant on a given surface element.
These charges are included in a column vector $\mathbf{h}$. This vector can be computed from a matrix-vector multiplication
\begin{equation}
 \mathbf{h}=\textsc{A}^{-1}\mathbf{c},
\end{equation}
where vector $\mathbf{c}$ contains the normal components of the electric field in the centers of the tiles and the matrix $\textsc{A}$ depends on the geometry of the dielectric boundary. Filling and inverting the matrix is a very time consuming process. In our simulations for ion channels\ \cite{boda-jcp-125-2006,boda-prl-98,boda-bj-93-2007} we used the fact that the dielectric boundary at the surface of the protein does not change during the simulation. Thus, the matrix can be precalculated at the beginning of the simulation and it does not really contribute to simulation time. The matrix-vector multiplication is also a time consuming step, but the simulations are still feasible. In the case of an ER fluid, on the contrary, the particles are moving during the simulation, the geometry of the dielectric boundary is constantly changing, and the matrix should be filled and inverted in every simulation step. This is why simulation is technically impossible using the DS model.

\section{The interaction energy between two dielectric spheres in an electric field}

Let us consider two DSs at a distance $r_{12}$ from each other. The dielectric constant inside sphere 1 (S$_{1}$) is $\epsilon_{1}$ and its radius is $a_{1}$. Similarly, the dielectric constant inside sphere 2 (S$_{2}$) is $\epsilon_{2}$ and its radius is $a_{2}$. In our calculations, we use spheres of equal unit radii $a_{1}=a_{2}=a=1$. The spheres are embedded in a dielectric $\epsilon_{\mathrm{w}}$. A homogeneous external electric field $\mathbf{E}$ is exerted on the system with strength of unity $E=1$, so the only free charges in the system are the electrode charges $\sigma$ and $-\sigma$ that raise this electric field. The potential of this field is $\Psi_{q}(\mathbf{r})=-\textbf{E}\cdot \mathbf{r}$.

The total energy of the system is $W=W_{q}+W_{h}$, where $W_{q}$ is the free charge --- free charge interaction energy, while $W_{h}$ is the free charge --- induced charge interaction energy. The latter, by symmetry, can be computed as the energy of the induced charges in the potential field of the free charges:
\begin{equation}
 W_{h}=\dfrac{1}{2} \int h(\mathbf{r}) \Psi_{q}(\mathbf{r}) \rd V = -\dfrac{1}{2} \int h(\mathbf{r}) (\textbf{E}\cdot \mathbf{r}) \rd V.
\end{equation}
Since the external field is constant, $\mathbf{E}$ can be brought out from the integral and the energy becomes
\begin{equation}
 W_{h}= - \dfrac{1}{2} \textbf{E}\cdot  \int h(\mathbf{r})\mathbf{r} \rd V.
\end{equation}
The integral can be divided into two integrals on the two spheres
\begin{equation}
  \int h(\mathbf{r})\mathbf{r} \rd V=  \int_{\mbox{S}_{1}} h(\mathbf{r})\mathbf{r} \rd V+  \int_{\mbox{S}_{2}} h(\mathbf{r})\mathbf{r} \rd V= \mathbf{p}_{1}+\mathbf{p}_{2}\,,
\end{equation}
where $\mathbf{p}_{i}$ is the dipole moment on the S$_{i}$ sphere.
So the energy is
\begin{equation}
 W_{h}=-\dfrac{1}{2}\left( \mathbf{E}\cdot \mathbf{p}_{1} + \mathbf{E}\cdot \mathbf{p}_{2} \right) = -\dfrac{1}{2}E\left( p_{1} +p_{2} \right)  ,
\end{equation}
where we defined the electric field as pointing in the direction of the $z$-axis $\mathbf{E}=E\mathbf{k}$ and $p_{i}$ is the $z$-component of the dipole moment.
Of course, the dipole moment depends on the mutual position $\mathbf{r}_{12}=\mathbf{r}_{1}-\mathbf{r}_{2}$ (with respect to the electric field) of the particles: $p_{i}=p_{i}(\mathbf{r}_{12})$.

The interparticle potential energy between the two spheres is defined as the difference between the total energies for a given mutual position and for the case when the spheres are infinitely far from each other:
\begin{equation}
\phi_{\mathrm{DS}}(\mathbf{r}_{12}) = W_{h}(\mathbf{r}_{12})-W_{h}(\infty).
\label{DS}
\end{equation}
Realize that $W_{q}$ drops out of this equation because it does not depend on the mutual position of the spheres. Since the dipole moment for the isolated sphere is $\mathbf{p}_{i}=\alpha_{i}\mathbf{E}$  (when the two spheres are infinitely far away), the self-energy is
\begin{equation}
W_{h}(\infty) = -\dfrac{1}{2}E^{2}(\alpha_{1}+\alpha_{2}) .
\end{equation}

\section{The point dipole models}

The interaction potential between two point dipoles is
\begin{equation}
 \phi_{\mathrm{DD}}(\mathbf{r}_{12},\mathbf{p}_{1},\mathbf{p}_{2})=
-3 \dfrac{ (\mathbf{p}_{1}\cdot \mathbf{r}_{12})(\mathbf{p}_{2}\cdot \mathbf{r}_{12})}{r_{12}^{5}}
 + \dfrac{\mathbf{p}_{1}\cdot \mathbf{p}_{2}}{r_{12}^{3}} \,,
\label{phidd}
\end{equation}
where $\mathbf{p}_{i}$ is the point dipole moment in the center of sphere S$_{i}$ [given by equation~\eqref{dipmom}, therefore, $\mathbf{p}_{i}$ is also a result of polarization] and $r_{12}=|\mathbf{r}_{12}|$.
Since $\mathbf{p}_{1}$ and $\mathbf{p}_{2}$ are parallel to each other and to $\mathbf{E}$, this potential can be written as follows:
\begin{equation}
 \phi_{\mathrm{DD}}(r_{12},\cos \theta, p_{1},p_{2})=
\dfrac{ p_{1} p_{2}}{r_{12}^{3}} \left(-3 \cos^{2}\theta +1 \right),
\label{uddspec}
\end{equation}
where $\theta$ is the angle between the vectors $\mathbf{r}_{12}$ and $\mathbf{p}_{i}$.

In the above potential, the dipole moments induced by the external fields on individual isolated dielectric spheres were used. When the spheres are close to each other, nevertheless, not only the external electric field, but also the electric field produced by the dipole of the other particle acts on this sphere. This effect can be taken into account with the polarizable point dipole model, where, in addition to the permanent dipole induced by the external field, an induced dipole appears
\begin{align}
\mathbf{p}_{1}^{\mathrm{ind}}(\mathbf{E}_{2})&=\alpha_{1}\mathbf{E}_{2}\,, \nonumber \\
\mathbf{p}_{2}^{\mathrm{ind}}(\mathbf{E}_{1})&=\alpha_{2}\mathbf{E}_{1} \,,
\label{palfaE}
\end{align}
where the electric field $\mathbf{E}_{i}$ is the electric field produced by the dipoles $\mathbf{p}_{i}$ and $\mathbf{p}_{i}^{\mathrm{ind}}$ at the position of the other dipole:
\begin{align}
\mathbf{E}_{1}(\mathbf{p}_{1},\mathbf{p}^{\mathrm{ind}}_{1})&=
3 \dfrac{\left[ (\mathbf{p}_{1} + \mathbf{p}_{1}^{\mathrm{ind}})\cdot \mathbf{r}_{12}\right]   \mathbf{r}_{12}}{r_{12}^{5}}-
       \dfrac{\mathbf{p}_{1}+\mathbf{p}_{1}^{\mathrm{ind}}}{r_{12}^{3}} \,,
\nonumber \\
\mathbf{E}_{2}(\mathbf{p}_{2},\mathbf{p}^{\mathrm{ind}}_{2})&=
3 \dfrac{\left[ (\mathbf{p}_{2} + \mathbf{p}_{2}^{\mathrm{ind}})\cdot \mathbf{r}_{12}\right]   \mathbf{r}_{12}}{r_{12}^{5}}-
       \dfrac{\mathbf{p}_{2}+\mathbf{p}_{2}^{\mathrm{ind}}}{r_{12}^{3}} \,.
\label{Efromp}
\end{align}
Since the induced dipoles produce fields that, in turn, induce dipoles, equations~\eqref{palfaE} and \eqref{Efromp} have to be solved iteratively \cite{vesely}.
After the induced dipoles are obtained, the interaction between $\mathbf{p}_{1}$ and $\mathbf{p}^{\mathrm{ind}}_{2}$ as well as the interaction between $\mathbf{p}_{2}$ and $\mathbf{p}^{\mathrm{ind}}_{1}$  are computed using equation~\eqref{phidd}. Then, we add this correction to $\phi_{\mathrm{DD}}(\mathbf{r}_{12},\mathbf{p}_{1},\mathbf{p}_{2})$ thus obtaining $\phi_{\mathrm{PD}}(\mathbf{r}_{12},\mathbf{p}_{1},\mathbf{p}_{1}^{\mathrm{ind}},\mathbf{p}_{2},\mathbf{p}_{2}^{\mathrm{ind}})$, where PD stands for ``polarizable dipole''.

\section{Results and discussion}

We consider two cases that we call \textit{parallel} and \textit{antiparallel} cases.
In the \textit{parallel} case, both spheres have dielectric constants smaller than that of the surrounding medium. Thus, the dipoles induced on the spheres point into the same direction. We use the value $\epsilon_{1}=\epsilon_{2}=2$ in this study. The dielectric constant of the solvent is $\epsilon_{\mathrm{w}}=4$. In the \textit{antiparallel} case, the dielectric constant in one sphere is larger than $\epsilon_{\mathrm{w}}$, while the dielectric constant in the other sphere is smaller than $\epsilon_{\mathrm{w}}$. Thus, the dipoles induced by $\mathbf{E}$ on the two spheres point into opposite directions.  We use the values $\epsilon_{1}=6$ and $\epsilon_{2}=2$ in this study.

\begin{figure}[!h]
\begin{center}
\scalebox{0.5}{\includegraphics*{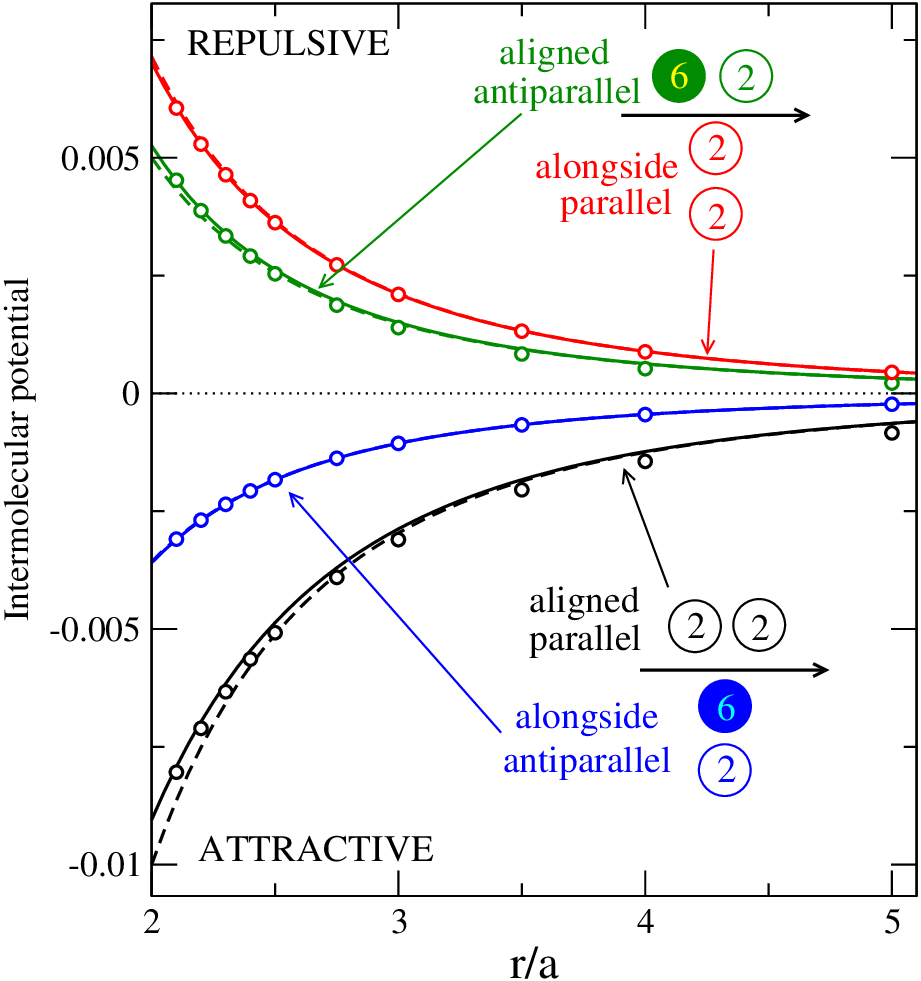}}
\end{center}\caption{(Color online) The interaction energy between two spheres (\textit{parallel} case: $\epsilon_{1}=\epsilon_{2}=2$, and \textit{antiparallel} case: $\epsilon_{1}=6$ and $\epsilon_{2}=2$) in two different positions: (1) the electric field is \textit{parallel} to the line connecting the centers of the spheres (\textit{aligned} case)  and (2) the electric field is perpendicular to the line connecting the centers of the spheres (\textit{alongside} case). The energy is computed from the three various models: symbols: the DS  model, solid line: the PD model, and dashed line: the DD model.}
\label{Fig1}
\end{figure}
Figure~\ref{Fig1} shows the interaction energies as a function of the distance between the two spheres (this distance will be denoted by $r=r_{12}$ henceforth) for various situations. Based on the dielectric constants of the spheres, we can consider the \textit{parallel} ($\epsilon_{1}=\epsilon_{2}=2$) and the \textit{antiparallel} ($\epsilon_{1}=2$ and $\epsilon_{2}=6$) cases. Based on the mutual position of the spheres, we can consider the \textit{aligned} ($\mathbf{E}\parallel \mathbf{r}_{12}$) and the \textit{alongside} ($\mathbf{E}\perp \mathbf{r}_{12}$) positions. As seen in figure~\ref{Fig1}, the \textit{aligned} position is the low-energy position for the \textit{parallel} case (the classical head-to-tail situation), while the \textit{alongside} position is the low-energy position for the \textit{antiparallel} case (negative, attractive interaction energies). The interactions are repulsive for the other cases (\textit{aligned antiparallel} and \textit{alongside parallel}). All these energies are larger in magnitude when the particles are closer to each other.


The agreement between the DD and DS potentials is surprisingly good (dashed lines vs. symbols in figure~\ref{Fig1}). Some deviation occurs for small interparticle distances because the DD approximation is not satisfactory when the two charge distributions are close to each other. The mutual polarization of the spheres can be taken into account by the PD model (solid lines in figure~\ref{Fig1}). The agreement with the DS results is excellent.

These results imply that the DD potential is an appropriate model of ER fluids in computer simulations.
The good agreement is, nonetheless, surprising and counter-intuitive.
It is not obvious from the first glance that the potential in equation~\eqref{DS} agrees with the potential in equation~\eqref{phidd}.
To shed light on this, let us consider that the dipole moment induced on a sphere S$_{i}$ is
\begin{equation}
p_{i}=\int_{\mbox{S}_{i}} \left[ h_{0,i}(\mathbf{r}) + \Delta h_{i}(\mathbf{r}) \right]  z \rd a,
\label{eq17}
\end{equation}
where $h_{0,i}(\mathbf{r})$ is the distribution on the isolated sphere and $\Delta h_{i}(\mathbf{r})$ is the distortion resulting from the effect of the other particle. The energy is obtained by multiplying by $-\frac{1}{2}E$. The term containing $h_{0,i}(\mathbf{r})$ is canceled by $W_{h}(\infty)$, so the interaction energy is simply
\begin{equation}
W_{h}=-\dfrac{1}{2}E \left[ \int_{\mbox{S}_{1}} \Delta h_{1}(\mathbf{r}) z \rd a + \int_{\mbox{S}_{2}} \Delta h_{2}(\mathbf{r}) z \rd a \right] .
\label{eq18}
\end{equation}
The distortion of the induced charge distribution on sphere S$_{1}$, for example, is proportional to the induced charge on this sphere:
more induced charge can be distorted more ($\Delta h_{1}\propto h_{1}\propto p_{1}$).
It is also proportional to the induced charge on the other sphere, because the electric field of sphere 2 that polarizes sphere 1 linearly depends  on $p_{2}$. This electric field depends on the cube of the distance inversely ($\propto r^{-3}$).
Finally, it will depend on the mutual position of the spheres which means an angle-dependent factor.
In summary,
\begin{equation}
W_{h} \propto p_{1} p_{2} r^{-3} f(\theta),
\end{equation}
which is exactly the form of the dipole-dipole potential in equation~\eqref{uddspec} for the special case of parallel dipoles.
\begin{figure}[h]
\begin{center}
\scalebox{0.5}{\includegraphics*{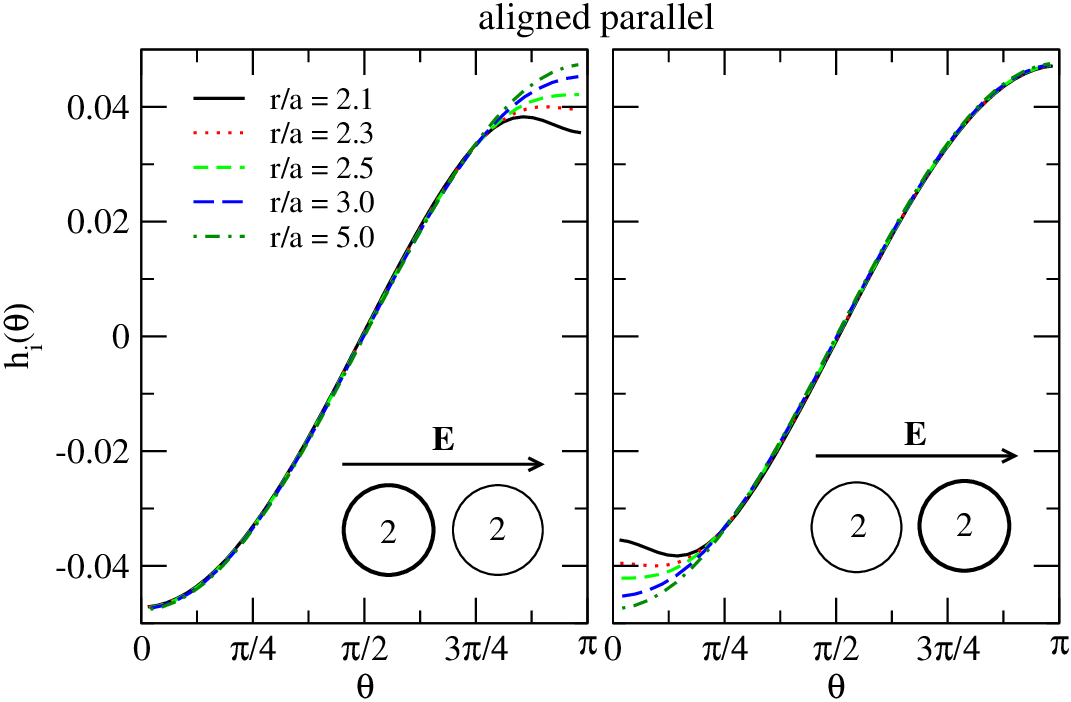}}
\end{center}\caption{(Color online) The induced charge on the surface of the spheres for different interparticle distances ($r/a$) for the \textit{aligned} \textit{parallel} case ($\epsilon_{1}=\epsilon_{2}=2$). The angle $\theta$ is closed by the vector pointing from the sphere center to the point on the surface and the vector of the electric field. Left hand panel shows the induced charge for the left hand side sphere, while the right hand panel shows the induced charge for the right hand side sphere. The points of the two spheres that are in the closest proximity correspond to $\theta=\pi$ for the left hand sphere and $\theta =0$ for the right hand sphere. These are the regions where the induced charge is the most distorted when $r/a$ is small.}
\label{Fig2}
\end{figure}
\begin{figure}[!h]
\begin{center}
\scalebox{0.5}{\includegraphics*{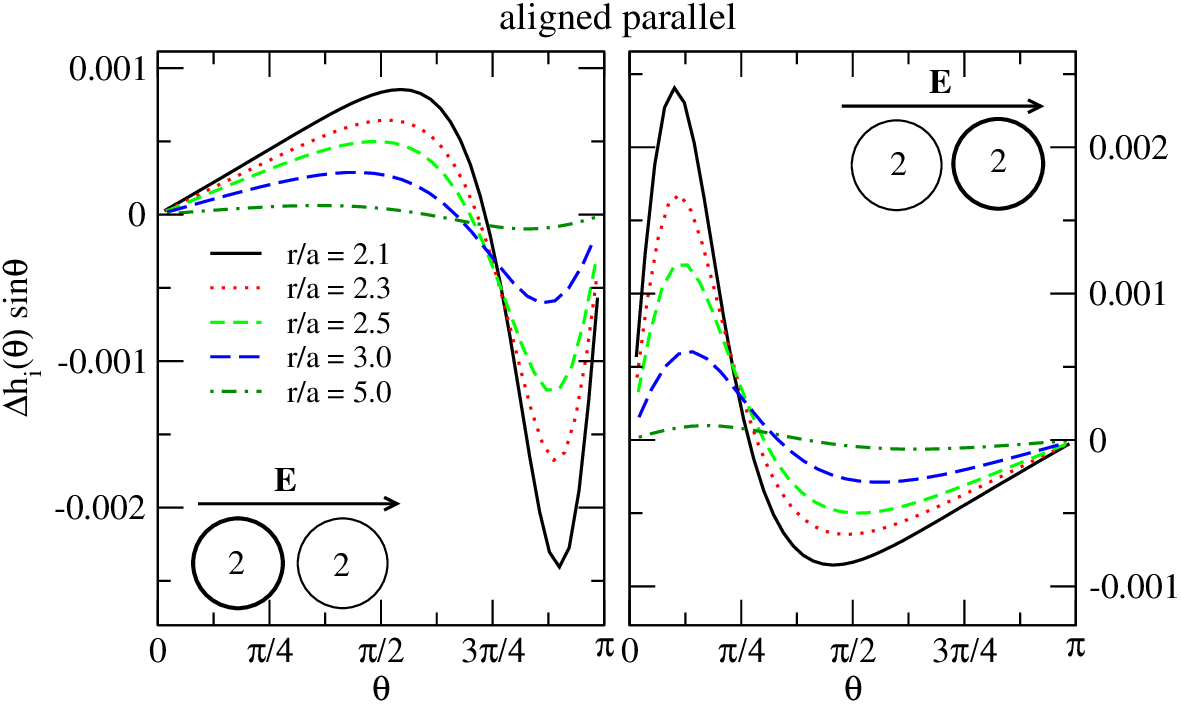}}
\end{center}\caption{(Color online) The distortion of the induced charge shown in figure~\ref{Fig2} computed from $\Delta h_{i}(\theta) = h_{i}(\theta) - h_{0,i}(\theta) = h_{i}(\theta) -\frac{3}{4\pi} p_{i}\sin \theta$.}
\label{Fig3}
\end{figure}

The mutual polarization of the two spheres is illustrated in figure~\ref{Fig2} for the \textit{aligned parallel} case.
The induced charges on the surfaces of the spheres are plotted as functions of the angle with the electric field for various interparticle distances.
For large distances, the charge distribution is symmetrical and the dipole moments on both of the spheres are the values for isolated spheres $\alpha_{i}=(2-4)/(2+2\times4)=-0.2$ (for $E=1$). The zero value of $\theta$ corresponds to the direction of the field, so the induced charge is negative on the hemisphere in the direction of the field and positive on the opposite hemisphere. This corresponds to a dipole moment whose direction is opposite to that of the field in agreement with the negative value of the polarizability.

When the distance between the spheres decreases, the symmetrical charge distribution is distorted.
The charge distribution of sphere S$_{2}$ pulls some extra negative charge on the tip of sphere S$_{1}$ that is in the closest proximity to sphere S$_{2}$. This extra negative charge is taken from all over the surface of the sphere, so the opposite positive charge appears there but on a larger surface, and this deviation in surface density is hardly distinguishable on the scale of  figure~\ref{Fig2}.

The distortion of the $h_{i}(\mathbf{r})$ surface charge is exactly the $\Delta h_{i}(\mathbf{r})$ surface charge introduced in equation~\eqref{eq17} and shown in figure~\ref{Fig3} for the case considered in figure~\ref{Fig2}.
The $\Delta h_{i}(\theta)$ function is multiplied by $\sin \theta$, so the total charge on a ring at angle $\theta$ is plotted. It is seen now that the integral of this charge distribution is zero.

The interaction energy of this charge with the external field [see equation~\eqref{eq18}] produces almost the same energy as that computed from the DD interaction potential [see equation~\eqref{phidd}] using the point dipoles corresponding to the undistorted charge distributions, $h_{0,i}(\mathbf{r})$.

The negative value means that the dipole is directed opposite to the electric field. Therefore, the total energy in this case is 0.2. When the spheres are close to each other, the charge distributions are slightly distorted on them due to polarization. This distortion is larger where the two spheres are in the closest proximity to each other.  This corresponds to a slightly different dipole moment and a slightly different energy. The difference between this energy and 0.2 gives the $\phi_{\mathrm{DS}}(\mathbf{r}_{12})$ interaction energy. This energy is very close to the $\phi_{\mathrm{DD}}$ interaction energy between the dipoles, which is quite surprising given the fact that the $\phi_{\mathrm{DS}}$ energy is the result of a distortion of the induced charges and thus, is the result of a change of the dipoles, while  $\phi_{\mathrm{DD}}$ is computed from the dipole moments fixed at their values at infinite separation.

\begin{figure}[h]
\begin{center}
\scalebox{0.5}{\includegraphics*{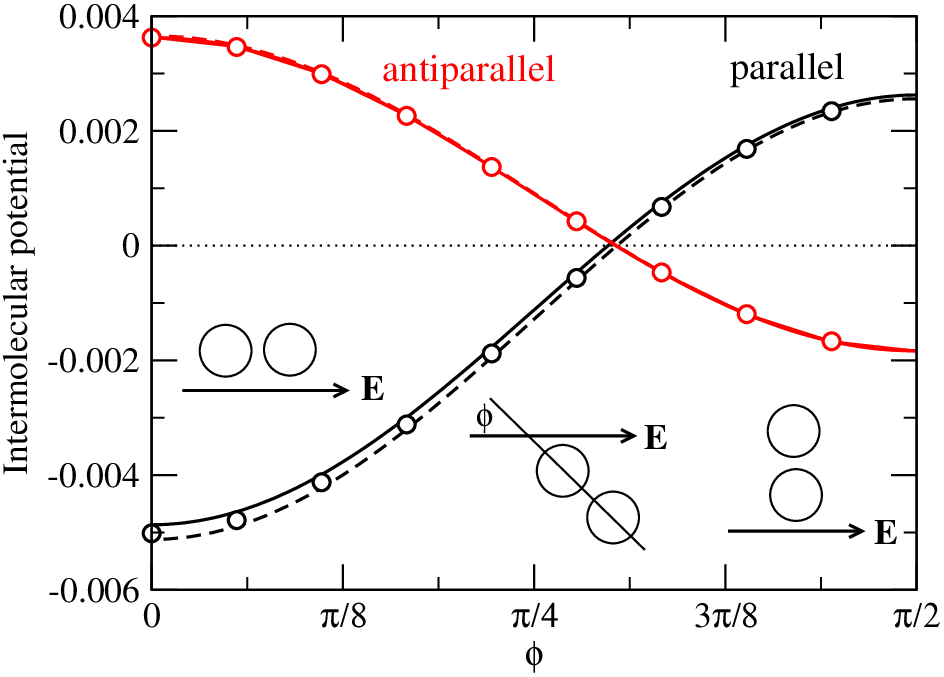}}
\end{center}\caption{(Color online) The interaction energy between two spheres for the \textit{parallel} and \textit{antiparallel} cases as a function of the angle between the electric field and the vector connecting sphere centers for $r/a=2.5$. The energy is computed from the three models. The meaning of curves and symbols is the same as in figure~\ref{Fig1}.}
\label{Fig4}
\end{figure}

In the next step, we study the effect of the mutual angular position of the spheres for a given distance: we fix $r/a=2.5$ and change the angle between $\mathbf{r}_{12}$ and $\mathbf{E}$ from 0 to $\pi /2$. Figure~\ref{Fig4}  shows the results as obtained from various models. Similar conclusions can be drawn as in the case of figure 1 for
the comparison of the various methods.
For the \textit{parallel} case, the angle $\theta =0$ corresponds to the minimum energy head-to-tail position. The angle $\theta =\pi/2$ corresponds to the maximum energy position where the dipoles of the spheres are next to each other pointing to the same direction. This means that the chains will repulse each other if the particles are in the same planes. A shifted position of chains corresponds to a more stable configuration at high densities when the chains are forced to be close to each other.
\begin{figure}[!h]
\begin{center}
\scalebox{0.5}{\includegraphics*{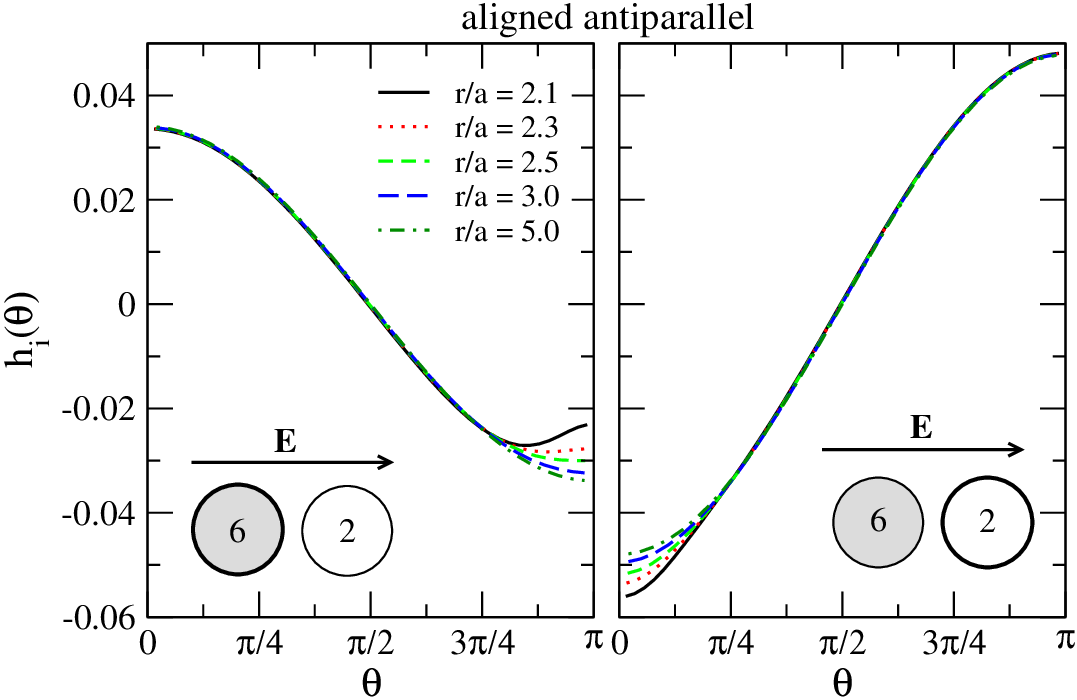}}
\end{center}\caption{(Color online) The induced charge on the surface of the spheres for different interparticle distances ($r/a$) for the \textit{aligned} \textit{antiparallel} case ($\epsilon_{1}=6$ and $\epsilon_{2}=2$). The angle $\theta$ is closed by the vector pointing from the sphere center to the point on the surface and the vector of the electric field. Left panel shows the induced charge for the left hand side sphere, while the right panel shows the induced charge for the right hand side sphere. The points of the two spheres that are in the closest proximity correspond to $\theta=\pi$ for the left sphere and $\theta =0$ for the right sphere. These are the regions where the induced charge is the most distorted when $r/a$ is small.}
\label{Fig5}
\end{figure}

For large distances, the charge distributions are symmetrical, which corresponds to point dipoles in the centers of the spheres with dipole moments $\alpha_{1}=(6-4)/(6+2\times4)=0.143$ and $\alpha_{2}=(2-4)/(2+2\times4)=-0.2$ (with $E=1$).

In the \textit{antiparallel} case, similar conclusions can be drawn except
that the configurations for the minimum and the maximum energy are interchanged. Here the minimum energy position is when the two spheres are next to each other with dipoles directed in opposite directions (blue symbols and curves in figure~\ref{Fig1} and $\theta =\pi/2$ in figure~\ref{Fig4}). The maximum energy position is when the two dipoles are aligned on the same line, a ``head-to-head'' position (green symbols and curves in figure~\ref{Fig1} and $\theta =0$ in figure~\ref{Fig4}). The induced charge is shown in figure~\ref{Fig5}.
The profiles are different for the two spheres because the polarizabilities of the two spheres are different now;
different both in sign and magnitude. The profiles for the right hand sphere ($\epsilon_{2}=2$) are similar to those in figure~\ref{Fig2}, while the profiles for the left hand sphere ($\epsilon_{1}=6$) have decreasing tendency as a function of $\theta$.
The polarizability of this sphere is positive and smaller in magnitude
than the polarizability of the other sphere: $|\alpha_{1}|=0.143$, while $|\alpha_{2}|=0.2$.

\section{Summary}

We have presented calculations for the interaction potential between two electrorheological particles, which, in a more detailed description, can be modelled as DSs immersed in a continuum dielectric that has a dielectric constant different from that of the sphere. We have shown that the interaction energy originated from the distortion of the induced charge distribution on the surface of the sphere as an effect of the presence of the other sphere is well reproduced by point dipoles placed in the centers of the spheres. Surprisingly, even the DD model (where this dipole is fixed at the value of the isolated sphere) gives a reasonable description.

We conclude that the DD or the PD models are useful simplified representations of the DS model for application in computer simulations. The potential acting between ER particles is used to calculate the energies in Monte Carlo simulations. Forces used in molecular dynamics or Brownian dynamics simulations can be straightforwardly derived from the potentials.

\section*{Acknowledgement}

We acknowledge the support of the Hungarian National Research Fund (OTKA K68641) and the J\'anos Bolyai
Research Fellowship.
The present publication was realized with the support of the project T\'AMOP--4.2.2/A--11/1/KONV--2012--0071 and T\'AMOP--4.1.1/C--12/1/KONV--2012--0017.

\ukrainianpart

\title{Походження міжчастинкового потенціалу електрореологічних плинів}

\author{Д. Бода\refaddr{label1}, М. Валіско\refaddr{label1}, І. Салаі\refaddr{label2}}

\addresses{\addr{label1} Факультет фізичної хімії, університет Паннонії, Веспрем, Угорщина
\addr{label2} Інститут фізики і мехатроніки, університет Паннонії, Веспрем, Угорщина}

\makeukrtitle

\begin{abstract}
Частинки електрореологічних плинів можуть бути змодельовані як
діелектрична сфера, занурена в діелектричне середовище. Коли
прикласти  зовнішнє поле, на поверхнях сфер індукуються
поляризаційні заряди, які можна представити як точкові диполі,
розміщені в центрах сфер.  Коли дипольні сфери є близько одна до
одної, розподіли індукованих зарядів є спотворені електричним полем
сусідніх дипольних сфер. Таким є походження  потенціалу взаємодії
між дипольними сферами. Обчислення цієї енергії займає багато часу,
тому модель дипольних сфер не може бути використана в молекулярних
симуляціях. У цій статті ми показуємо, що взаємодія між точковими
диполями  належним чином апроксимує взаємодію дипольних сфер. Модель
поляризованих точкових диполів забезпечує кращі результати, але ця
модель не є попарно адитивною, отже не є практичною в симуляціях
частинок.
\keywords електрореологічні плини, симуляція, міжчастинковий
потенціал
\end{abstract}

\end{document}